\documentclass[aps,prb,twocolumn,amsmath,floatfix]{revtex4-2}
\usepackage[x11names]{xcolor}
\usepackage[pdftex]{graphicx}                    
\usepackage{dcolumn}   
\usepackage{bm}                                                      
\usepackage{amssymb}
\usepackage{ulem}

\newcommand{\bphi}{\mbox{\boldmath $ \phi $}}

\begin{document}

\title{Defect-induced spin textures in magnetic solids}

\author{M. E. Zhitomirsky}
\affiliation{Universit\'e Grenoble Alpes, CEA, IRIG, PHELIQS, 38000 Grenoble, France}
\affiliation{Institut Laue-Langevin, 71 avenue des Martyrs, 38042 Grenoble, France}

\author{Vijay B. Shenoy}
\affiliation{
Centre for Condensed Matter Theory, Department of Physics, Indian Institute of Science, 
Bangalore 560 012, India}

\author{Roderich Moessner}
\affiliation{Max-Planck-Institut f\"ur Physik Komplexer Systeme, 
N\"othnitzer Stra\ss e 38, D-01187 Dresden, Germany}

\date{\today}

\begin{abstract}
Vacancy defects in isotropic noncollinear antiferromagnets  produce long-range spin textures.
By developing a ``magnetic elasticity theory'', we  demonstrate that a vacancy-induced 
readjustment in the spin configuration decays algebraically with distance. 
The power law exponent depends on the multipole moment of a local spin deformation, which in turn is 
determined by the lattice symmetry and an equilibrium spin configuration in the absence of defects. 
The role of these two factors is highlighted for the $J_1$--$J_2$ Heisenberg model on a kagome lattice.
A vacancy in this model generates spin deformations that decay as $1/r^2$ for the $q=0$ ground 
state and as a $1/r$ for the $\sqrt{3}\times\sqrt{3}$ magnetic structure.  The analytic conclusions are 
confirmed by extensive numerical simulations. We also compute the fractional magnetic moments 
associated with vacancies and other lattice defects. Our results shed light on relative fragility of 
different magnetic structures with respect to spin glass formation at higher doping levels.
\end{abstract}

\maketitle

\section{Introduction}

The role of structural disorder is a recurring topic in experimental studies of magnetic materials, 
for recent examples see
\cite{Bobroff09,Dutton10,Okuda11,Simutis13,Kim14,Niven14,Ross16,Smirnov17,Orlova18,Mirebeau18,Bowman19,%
Lim20,Facheris20,Pughe23,Goodge23,Chandana24}.
It is now widely recognized that  disorder is not only a source of unwanted 
or parasitic effects but may also produce interesting physics and enable
various applications \cite{Jungwirth14,Dietl15,Morin16}.
Early theoretical studies on  defect-induced phenomena in magnetically ordered crystals 
date back more than half a century, see the articles by de Gennes \cite{DeGennes59}, Villain \cite{Villain79} and
references therein. The discovery of  high-temperature superconductivity in doped antiferromagnets 
gave a new impetus to  theoretical studies 
\cite{Aharony88,Parker88,Vannimenus89,Glazman90,Aristov90,Gawiec91}. 
More recently, the effect of structural disorder was investigated for geometrically 
frustrated magnets and related systems with extra `accidental' degeneracies among classical ground 
states. Random impurities and bond defects can lower the energy for certain ordered spin configurations providing an example of the order by disorder selection
\cite{Henley89,Long89,Slon91,Fyodorov91,Savary11,Maryasin13,Maryasin14,Andreanov15}.
Alternatively, if the classical degeneracy is high, randomness in the magnetic Hamiltonian  may 
induce either a spin glass state \cite{Bellier01,Saunders07,Andreanov10,Andrade18,Dey20}
or various kinds of quantum and classical spin liquids \cite{Henley01,Rehn15,Savary17,Bilit17,Kimchi18,Consoli24}.
In addition, considerable attention was devoted to properties of an isolated nonmagnetic impurity 
(vacancy) in ordered and disordered antiferromagnets
\cite{Sigrist96,Schiffer97,MoessnerBerlinsky99,Sachdev99,Sachdev03,Hoglund07,Eggert07,Sen11,Wollny11,Weber12,%
Sen12,Maryasin15,Utesov15,Lin16,Hayami19}. At low temperatures,  vacancies can generate effective
paramagnetic moments, which have fractional values
\cite{Sen11,Wollny11,Weber12,Sen12,Maryasin15}.

A physical phenomenon common to many theoretical and experimental works is a local 
spin readjustment, or canting, in the vicinity of a structural defect. For magnets with 
spontaneously broken continuous symmetries, the resulting spin textures can extend over 
large distances. Such clouds of tilted spins 
surrounding nonmagnetic defects have a strong effect on various physical properties including
magnon transport and stability of magnetic structures 
with respect to transformation into a spin glass state upon increasing the density of defects.
Taking into account importance of spin currents for spintronics applications \cite{Maekawa13} 
and also a general interest in the role of defects in magnets, we revisit in our work the problem
of strain fields created by isolated defects in ordered magnetic structures.

Spin textures induced by defect exchange bonds are well understood by now    
\cite{Villain79,Aharony88,Parker88,Vannimenus89,Gawiec91}, see also Sec.~II. The associated strain field 
propagates with distance $r$ from the defect as $1/r^{D-1}$ both for two ($D=2$) and three ($D=3$) dimensional 
lattices. However, a consistent theoretical description of spin textures near a vacancy in noncollinear antiferromagnets 
is still lacking. A number of published papers have proposed different distance dependence, such as the  $1/r^D$ 
law \cite{Henley01,Weber12} or  the $1/r^{D+1}$ law \cite{Wollny11}, both of which supported by 
the numerical results. In essence, the available results imply that dimensionality alone is not sufficient to determine 
the asymptotic behavior of the strain field induced by a nonmagnetic vacancy.

We demonstrate below that the decay law also depends on {\it symmetry} of an unperturbed magnetic structure. We base our 
consideration on the elasticity approach to ordered magnetic states. This provides a simple analytic description for the spin canting 
far away from the defects. Our analytic predictions are confirmed by numerical results.
The rest of the paper is organized as follows. Section~\ref{CoplanarA} sets up a general framework and presents
 the analytical theory for the defect-induced spin textures in the coplanar magnetic states. The analytical results
are compared in Sec.~\ref{CoplanarN} with the numerically simulated spin textures for various two- and three-dimensional 
spin models with the coplanar ground states. Generalization of the analytical theory to the case of noncoplanar magnetic structures 
is given in Sec.~\ref{NoncoplanarA} and
the corresponding numerical results are included in Sec.~ \ref{NoncoplanarN}. 
Numerical results for the defect-induced moments produced by  spins vacancies are presented in  Sec.~\ref{Moments}. 
An overview of the obtained results and further discussion is provided in Sec.~\ref{Discussion}.

\section{Coplanar spin textures}
\label{Coplanar}

\subsection{Analytic theory}
\label{CoplanarA}

We consider the Heisenberg  model with classical unit length spins, $|{\bm S}_i|=1$,
\begin{equation}
\hat{\cal H} = \sum_{\langle ij\rangle} J_{ij}\, {\bm S}_i\cdot{\bm S}_j\,,
\label{H}
\end{equation}
where  $\langle ij\rangle$ denotes summation over bonds.
In a pure material without impurities, the Hamiltonian (\ref{H}) is translationally invariant and the exchange 
parameters $J_{ij}$ are functions of intersite distance ${\bm r}_{ij}$. Competing exchange interactions can stabilize various
noncollinear states \cite{Villain59,Yoshimori59}.
In this section we consider the case of {\it coplanar} magnetic structures at $T\to 0$.  
Without loss of generality, we assume that  the spins lie in  the $x$--$z$ plane and
 parameterize their orientation by the angle $\theta_i$:
\begin{equation}
{\bm S}^{0}_i = (\sin\theta_i , 0, \cos\theta_i) \,.
\label{S0}
\end{equation}
For spin models (\ref{H}) defined on a Bravais lattice, the classical ground state is typically a spin spiral
described by  propagation wavevector $\bm Q$ such that  $\theta_i = {\bm Q}\cdot {\bm r}_i$.
The wavevector $\bm Q$ corresponds to the minimum of
the Fourier transform of the exchange interactions
\begin{equation}
J_{\bm k} = \sum_j J_{ij}\,e^{i{\bm{kr}_{ij}}}.
\end{equation}
Impurities and other defects can modify locally the exchange interactions between spins. 
As a result, new perturbed exchange parameters $\tilde{J}_{ij}$ appear in the spin Hamiltonian
(\ref{H}). In the local picture of structural defects commonly used in literature, $\tilde{J}_{ij}\equiv J_{ij}$
beyond a certain distance from the defect.

The perturbed exchange bonds cause readjustment of spins producing a strain in 
the undistorted magnetic state (\ref{S0}). The early theoretical works have focused on the problem of a defect 
antiferromagnetic bond in collinear ferromagnets
\cite{Villain79,Aharony88,Parker88,Vannimenus89,Glazman90,Aristov90,Gawiec91}.
Once the strength of a perturbed bond exceeds a threshold value, a canted spin state arises in 
the vicinity of the defect, with a slowly decaying long-range tail. For a defect in a noncollinear 
antiferromagnet, spins begin to tilt for an arbitrarily weak modification of the perturbed bond(s).
The deformation field ${\bm m}_i$ can be defined as
\begin{equation}
{\bm m}_i = {\bm S}_i - {\bm S}^{0}_i \,,
\end{equation}
where ${\bm S}_i$ is an equilibrium orientation of spin  and ${\bm S}^{0}_i$ is its orientation the reference
state (\ref{S0}).  In such a case,  $|{\bm m}_i|\to 0$ at large distances for the defect. 

In order to obtain the equation for ${\bm m}_i$, we transform the spin Hamiltonian 
 to a local ``rotating'' frame associated with the undistorted reference state (\ref{S0}).
In this coordinate frame, $\hat{\bm z}_i$ axis is chosen along ${\bm S}^{0}_i$, 
$\hat{\bm y}_i = \hat{\bm y}$ points out of the spin plane, so that
$\hat{\bm x}_i =  \hat{\bm y}\times \hat{\bm z}_i$  is oriented within the plane. 
The  energy (\ref{H}) transforms to
\begin{eqnarray}
E({\bm m}) & = & \sum_{\langle ij\rangle} \tilde{J}_{ij} \bigl[ m_i^y m_j^y + \cos\theta_{ij}
\bigl( S_i^z S_j^z + m_i^x m_j^x\bigr) \nonumber \\
&&  \mbox{\quad\ } + \sin\theta_{ij} \bigl(   S_i^z m^x_j - m_i^x S_j^z \bigr) \bigr]\,, 
\label{Hrot}  
\end{eqnarray}
with $\theta_{ij} = \theta_i - \theta_j$ and  $S_i^z = \bigl(1- {m_i^x}^2 - {m_i^y}^2\bigr)^{1/2}$.

The linear terms in the energy functional (\ref{Hrot}) are explicitly given by
\begin{equation}
E_1({\bm m})  =  - \sum_{i} m_i^x h_i^x \,, \quad h_i^x = \sum_j \tilde{J}_{ij}\sin\theta_{ij}\,.
\label{E1}
\end{equation}
Here, $h_i^x$ is a transverse or canting field acting on a spin in the undistorted  state (\ref{S0}) 
in the presence of a structural defect. For sites connected to their neighbors by unperturbed bonds
$h_i\equiv 0$. A more general expression for the transverse field is
\begin{equation}
\bm{h}_i = \bm{S}^0_i \times \bigl(\bm{H}_i\times  \bm{S}^0_i\bigr), \quad
\bm{H}_i = - \sum_j \tilde{J}_{ij}\,  \bm{S}^0_j\,.
\end{equation} 
The above expression can be further simplified by noting that in the undistorted structure
spins are aligned with local fields:  $\bm{S}^0_i\parallel\bm{H}_i^0 = - \sum_j J_{ij}\bm{S}^0_j$. 
Accordingly, only distorted bonds
contribute to $\bm{h}_i$ and one can use an excess field $\tilde{\bm{H}}_i =  \bm{H}_i - \bm{H}_i^0$ in place of 
$\bm{H}_i$:
\begin{equation}
\bm{h}_i = \bm{S}^0_i \times \bigl(\tilde{\bm{H}}_i\times  \bm{S}^0_i\bigr), \quad
\tilde{\bm{H}}_i = - \sum_j \delta\tilde{J}_{ij}\,  \bm{S}^0_j\,.
\label{Htr}
\end{equation} 
Furthermore,   a  projection of the transverse field on the local $\hat{\bm x}_i$  axis is      
equal to the $y$-component of a torque:
\begin{equation}
h_i^x = \hat{\bm y} \cdot \bigl(\bm{S}^0_i\times \tilde{\bm{H}}_i \bigr) = \tau_i^y \,.
\label{Htrx}
\end{equation} 
Nonzero values of either $\bm{h}_i$ or  $\bm{\tau}_i$  signal  that a spin configuration is out of equilibrium. Description in terms of the transverse (canting) fields arises naturally in the context of planar magnetic structures, where $\bm{h}_i$ provides a direction of spin tilting on a site $i$ and serves as a source term  in the  linear equations on $m_i^x$ and $m_i^y$, see Eq.~(\ref{Green}). A  torque description is, however, more convenient for the noncoplanar states, Sec.~\ref{NoncoplanarA}.

Transverse fields on the defect sites produce a distribution of ${\bm m}_i$ away from the defect.
To proceed with an analytical description of the emerging spin textures, we assume that bonds are weakly perturbed and $|{\bm h}_i| \simeq  |\delta\tilde{J}_{ij}|  \ll |J_{ij}|$. 
In this case, it is sufficient to 
expand the energy functional  $E({\bm m})$ to the second order in small $m_i$. The corresponding strain energy is 
\begin{eqnarray}
E_2({\bm m}) & = & \sum_{\langle ij\rangle} J_{ij}\Bigl\{\! - \frac{1}{2} 
\cos\theta_{ij}\Bigl[(m_j^x\!-m_i^x)^2\! +(m_j^y\!-m_i^y)^2\Bigr]
\nonumber \\
&  &   \mbox{\qquad\quad} + m_i^y m_j^y\, (1-\cos\theta_{ij}) \Bigr\}\,,
\label{E2}
\end{eqnarray}
where it is sufficient to substitute for all exchange bonds their unperturbed values $J_{ij}$.
The subsequent calculations are performed for a spin 
spiral with $\theta_{ij} =\bm{Q}\cdot\bm{r}_{ij}$.

Minimization of  the quadratic form $E_1({\bm m})+E_2({\bm m})$ with respect to ${\bm m}_i$ yields a system of linear 
equations for two components of the deformation field:
\begin{eqnarray}
& & \sum_j J_{ij}\cos( {\bm Q}{\bm r}_{ij})\, m_j^x - J_{\bm Q}\, m_i^x = h_i^x \,,
\nonumber \\
& & \sum_j J_{ij} m_j^y - J_{\bm Q}\, m_i^y = h_i^y \,.
\label{Mlin}
\end{eqnarray}
For a planar magnetic state, defect bonds or vacancies generate no transverse field $h_i^y$, 
see Eqs.~(\ref{Hrot}). Still, we may consider a general perturbation with a finite $h_i^y$, 
which may arise, for example, because of a magnetic impurity.

The solution of Eq.~(\ref{Mlin})  can be constructed with the help of  the lattice Green's functions
\cite{Katsura71} defined in our case by
\begin{eqnarray}
& & \sum_j J_{ij}\cos( {\bm Q}{\bm r}_{ij})\, G_{xx}(\bm{r}_j) - J_{\bm Q}\, G_{xx}(\bm{r}_i) = \delta_{\bm{r}_i,0} \,,
\nonumber \\
& & \sum_j J_{ij} G_{yy}(\bm{r}_j) - J_{\bm Q}\, G_{yy}(\bm{r}_i) =  \delta_{\bm{r}_i,0} \,,
\end{eqnarray}
$\delta_{\bm{r},0}$ being the Kronecker symbol. In the momentum representation, the Green's
functions are given by
\begin{eqnarray}
&& G_{xx}^{-1}(\bm{k}) = \frac{1}{2}(J_{\bm{k}+\bm{Q}}+J_{\bm{k}-\bm{Q}})-J_{\bm Q}\,,
\nonumber \\
&& G_{yy}^{-1}(\bm{k}) =  {J_{\bm k}-J_{\bm Q}}\,.
\end{eqnarray}
Then, the solution of Eq.~(\ref{Mlin}) for an arbitrary distribution of canting fields $\bm{h}_i$ is expressed as
\begin{subequations}
\label{Green}
\begin{align}
m_i^x  & =  \frac{1}{N}\sum_{\bm k} \sum_j \frac{e^{i{\bm k} {\bm r}_{ij} }}
{\frac{1}{2}(J_{\bm{k}+\bm{Q}}+J_{\bm{k}-\bm{Q}})-J_{\bm Q}}\, h_j^x \ , 
\label{GreenX}
\\
m_i^y  & =  \frac{1}{N}\sum_{\bm k} \sum_j \frac{e^{i{\bm k} {\bm r}_{ij}}}
{J_{\bm k}-J_{\bm Q}}\, h_j^y \ ,
\label{GreenY}
\end{align}
\end{subequations}
where $N$ is the number of sites. At large distances, the momentum sums are dominated by regions, 
where denominators are small: ${\bm k}\to 0$ for Eq.~(\ref{GreenX}) and  ${\bm k}\to\pm{\bm Q}$ for Eq.~(\ref{GreenY}).
Performing a substitution   ${\bm k} = {\bm k} \pm{\bm Q}$ in Eq.~(\ref{GreenY}) and 
 expanding denominators in small ${\bm k}$, we obtain:
\begin{subequations}
\begin{align}
m_i^x  & \approx  \frac{1}{N}\sum_{\bm k} \: \frac{e^{i{\bm k}{\bm r}_i}}{\lambda |{\bm k}|^2}
\sum_j  h_j^x\, e^{-i{\bm k}{\bm r}_j}, 
\label{Mxa}  \\
m_i^y  & \approx e^{i{\bm Q}{\bm r}_i}   \frac{1}{N}\sum_{{\bm k}}\:  \frac{e^{i{\bm k}{\bm r}_i}}{\lambda|{\bm k}|^2}
\sum_j  h_j^y\, e^{-i(\bm{Q}+{\bm k}){\bm r}_j} + \textrm{c.\,c.}\,,
\label{Mya} 
\end{align}
\end{subequations}
where a positive constant  $\lambda$ is given by the second order  derivative  
$\lambda = \frac{1}{2}\nabla^2_{\bm{k}} J_{\bm k}$ at the ordering wave vector $\bm Q$. 
The main difference between in-plane
and out-of-plane responses consists in the staggered  $e^{\pm i{\bm Q}{\bm r}_i}$ prefactors for the later.
From now on, we will focus on the  in-plane response, $m_i^x$, which is relevant for structural defects,
and simplify notations: $h_i^x \to h_i$.

The in-plane distortion of spins $m_i^x$ caused by a lattice defect exhibits a 
power-law dependence at large distances. The asymptotic form of 
$m^x_i$ (\ref{Mxa}) follows from a solution of the Laplace's equation presented in  the
Fourier transformed form. Its real space behavior is determined by the small ${\bm k}$
expansion  of the transverse field source
\begin{equation}
\hat{Q}({\bm k}) = \sum_j h_j e^{-i{\bm k}{\bm r}_j} \,,
\label{Qsource}
\end{equation}
which is analogous to the multipole expansion for a system of point charges \cite{LL2}.

If a local perturbation is not weak, a full solution of the nonlinear problem based on the lattice energy 
(\ref{Hrot}) is required, see Sec.~\ref{CoplanarN}. Still, the quadratic approximation  is valid at 
large distances, where the spin tilting is small. Replacing $m^x_i$ with a continuous field 
$m^x({\bm r})$ and approximating
\begin{equation}
m_i^x - m_j^x \approx\, \bigr(\bm{r}_{ij}\!\cdot\!\bm{\nabla}\bigl)m^x(\bm{r})
\end{equation}
 one obtains from Eq.~(\ref{E2}) a gradient expansion for the strain energy
\begin{equation}
E_2 = \frac{1}{2} \int \textrm{d}^D r\, K_{\alpha\beta} (\nabla_\alpha m^x) \bigr(\nabla_\beta m^x\bigl)\,, 
\label{Egrad}
\end{equation}
where the elastic tensor  is expressed as
\begin{equation}
K_{\alpha\beta} = -\frac{1}{V_0} \sum_j J_{ij}\, r^\alpha_{ij}  r^\beta_{ij}  \cos(\bm{Qr}_{ij})\,,
\end{equation}
$V_0$ being the unit cell volume. If the ordering wave-vector $\bm{Q}$ corresponds to
 a high-symmetry point in the Brillouin zone, the tensor $K_{\alpha\beta}$ becomes diagonal and 
proportional to $\lambda$, see Eq.~(\ref{Mxa}). For a low symmetry, rescaling 
the space coordinates along the principal directions
of the second-rank tensor $K_{\alpha\beta}$ brings  (\ref{Egrad}) back to the isotropic form.

The in-plane spin readjustment  described by a tilting angle $\phi({\bm r})\approx m^x({\bm r})$  
satisfies the Laplace's equation 
\begin{equation}
\Delta \phi = 0  \,.
\label{DeltaMxy}
\end{equation} 
Its solution corresponds to an `electric' potential field produced by  a delta-like source term.  
The source term  can be still found from  Eqs.~(\ref{Htr}) and (\ref{Qsource}). The nonlinear corrections 
renormalize its amplitude, but leave the symmetry intact.

As an example, let us consider a defect bond that connects sites $l$ and $n$ with
a distinct exchange coupling $\tilde{J}$, which differs by  $\delta J = \tilde{J} - J$ from
the corresponding value in the surrounding lattice. The produced source term  
has a dipole symmetry, since the function $h_i$ in Eq.~(\ref{E1}) changes sign under 
the permutation $l\leftrightarrow n$. Hence, a bond defect generates a readjustment of spins
in the form of a dipole potential  \cite{Villain79}
\begin{equation}
 \phi({\bm r}) = \frac{{\bm d}\cdot{\bm r}}{|{\bm r}|^{D}} 
= \frac{d^\alpha\hat{r}^\alpha}{r^{D-1}}\ ,
\label{Dipole}
\end{equation}
where $\hat{\bm r} = {\bm r}/|{\bm r}|$.
The dipole moment ${\bm d}$ is parallel to the modified bond
${\bm d}\parallel{\bm r}_{i'j'}$ and in the linear approximation
$|{\bm d}|\sim |\delta J|$. 
In two and three dimensions, the defect induced  strains in the magnetic structure slowly decay with distance
as  $1/r$ and  $1/r^2$, respectively. Obviously, the power law behavior 
for $\phi({\bm r})$ is related
to the Goldstone mode for global rotations of the coplanar magnetic structure
about the orthogonal axis.

\begin{figure}[t]
\centerline{\includegraphics[width=0.99\columnwidth]{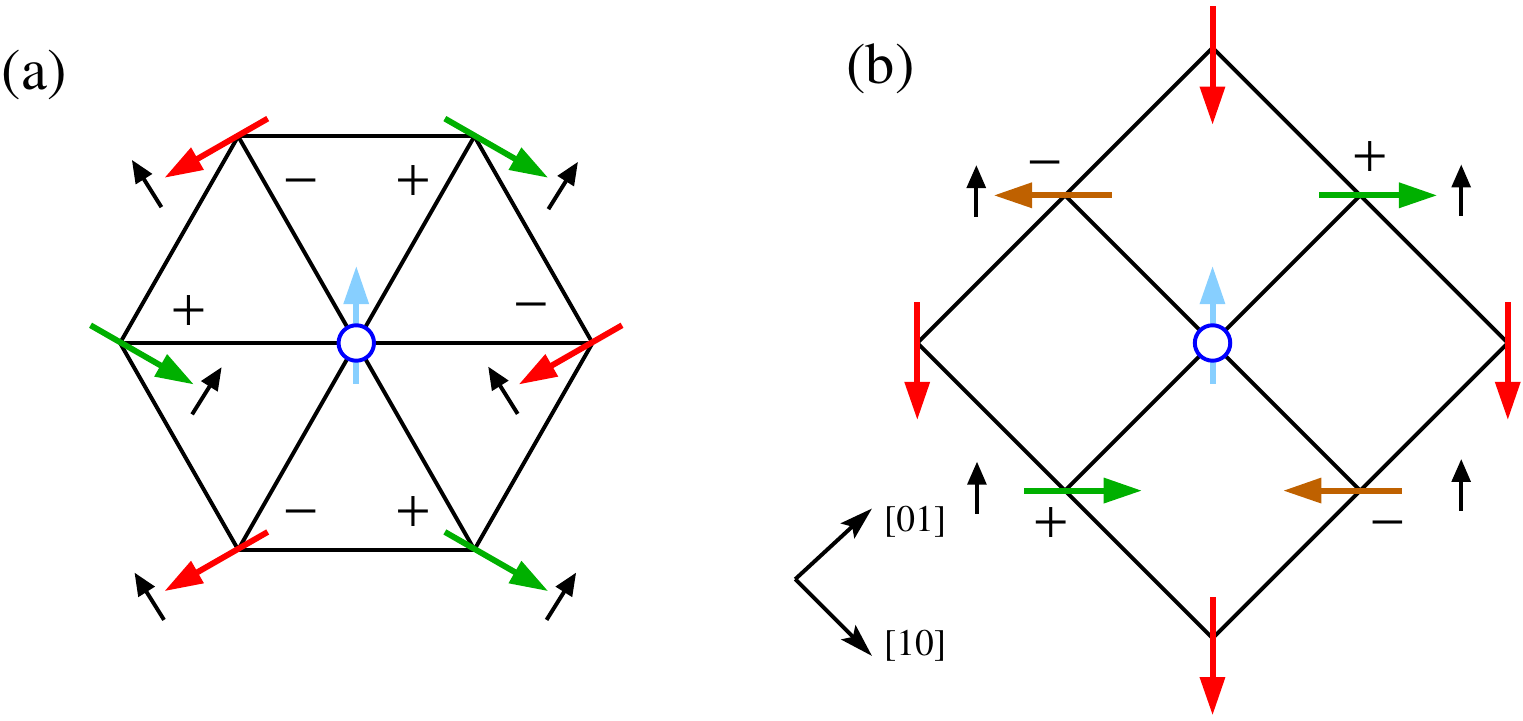}}
\vspace{5mm}
\centerline{\includegraphics[width=0.95\columnwidth]{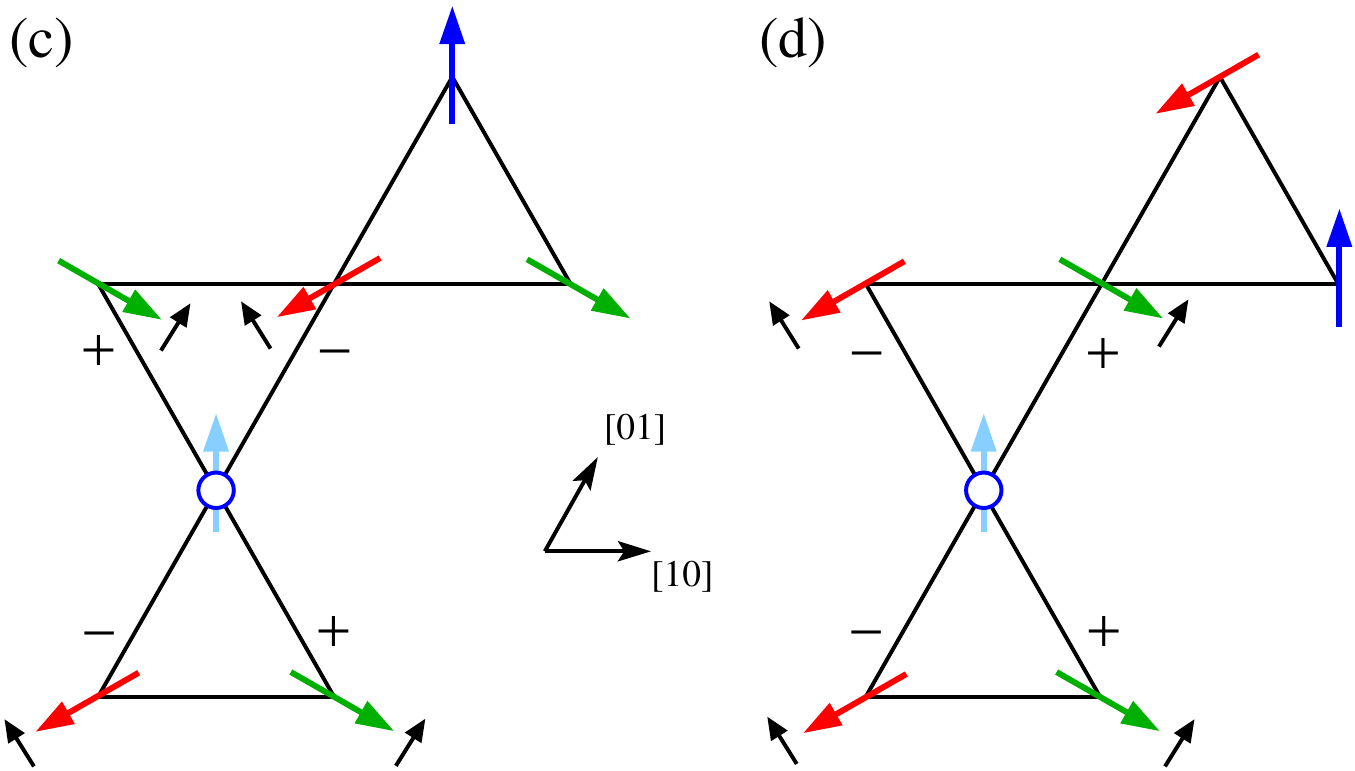}}
\caption{Distribution of local fields in  undistorted noncollinear structures around  a  vacancy (white circle). 
Transverse fields are shown by small black vectors.
The signs of $h_i^x$ in the rotated local frame (\ref{Htrx}) are indicated by  $+/-$ for 
spins in the nearest-neghbor shell.
(a) Triangular antiferromagnet; (b) $J_1$--$J_2$ square-lattice antiferromagnet with $J_2>0.5J_1$;
(c) the $q=0$ state for  $J_1$--$J_2$ kagome antiferromagnet 
($J_2>0$) and (d)  the $\sqrt{3}\times\sqrt{3}$ structure ($J_2<0$).}
\label{fig:Multipoles}
\end{figure}

We now proceed to examine the vacancy effect.  It is possible to model  a vacancy at a site $i=n$,
by switching off the surrounding bonds $\tilde{J}_{ni}\to 0$ instead of removing  the corresponding
spin $\bm{S}_n$. Therefore, the  above
theory  fully applies here.  For simple spin models such as  two-dimensional antiferromagnets
 with the nearest-neighbor exchanges,
the distribution of  transverse fields $h_i$ on sites adjacent to a vacancy can be obtained without any calculation 
using the relation $\phi_i \simeq h_i$. The signs of tilting angles $\phi_i$ can be easily drawn based on an observation
that spins on the  adjacent bonds become more antiparallel  to each other due to a partial release of frustration.
Figure~\ref{fig:Multipoles} 
illustrates the corresponding spin readjustments for four representative examples.
In more complex cases with several competing exchanges, one can compute
the transverse fields  $h_i$ directly  
from Eq.~(\ref{Htrx}) using the fact that  an excess field  $\tilde{\bm H}_i \simeq \bm{S}_n^0$ does not depend on $i$
for the nearest-neighbor spins  of site $n$ ($\delta{J}_{ni}=-J$).

We begin with the Heisenberg triangular antiferromagnet (TAF), Fig.~\ref{fig:Multipoles}(a). 
Due to geometric magnetic frustration, antiparallel alignment of spins cannot be simultaneously 
achieved for all nearest-neighbor bonds, and the lowest energy spin configuration corresponds 
to the $120^\circ$ magnetic structure. This magnetic structure is an example
of a magnetic spiral with the ordering wave vector ${\bm Q} = (4\pi/3,0)$, 
where the nearest-neighbor distance is used as a unit length. 

A vacant spin site relieves frustration in the surrounding triangular plaquettes and allows for 
the adjacent spins to align more antiparallel to each other. The associated spin 
tilting alternates around the vacant site as shown in Fig.~\ref{fig:Multipoles}(a) with 
$\phi_i, h_i \sim \cos 3\varphi$, where $\varphi$ is the azimuthal angle. Thus, the multipole expansion
for a canting-field source produced by a vacancy in TAF starts with an octupole term
$T^{\alpha\beta\gamma} = \sum_i h_i r_i^\alpha r_i^\beta r_i^\gamma$.
The spin texture with the octupole symmetry varies at large distances as \cite{LL2}:
\begin{equation}
\phi({\bm r}) \propto\frac{\textrm{Re}\,(\hat{x}+i\hat{y})^3}{r^{D+1}}
\label{Tabc}
\end{equation}
corresponding to the $1/r^3$ law in two dimensions.
A spin texture created around a vacancy in TAF possesses   three-fold rotation symmetry. 
In particular, spins do not tilt 
along the vertical line and two other equivalent crystallographic directions passing through the vacant site.
These directions correspond to zeros of the octupole $\cos 3\varphi$ harmonic 
and are further protected by the rotation symmetry.

The second model is the frustrated square lattice antiferromagnet (FSAF)
with nearest $J_1$ and next-nearest $J_2$ neighbor Heisenberg exchange interactions. 
For $J_2>\frac{1}{2} J_1$ the ordering wave vector is either ${\bm Q} = (\pi,0)$
or ${\bm Q} = (0,\pi)$ and the model has degenerate classical ground states. 
The degeneracy can be visualized as an arbitrary mutual orientation of two interpenetrating
square lattice antiferromagnets formed by the diagonal $J_2$ bonds.
At zero temperature, a vacancy enforces an orthogonal orientation  
of the two interpenetrating antiferromagnetic sublattices \cite{Henley89}. 
The resulting spin readjustments in the nearest-neighbor shell 
$\phi_i, h_i \sim \cos 2\varphi$, see 
Fig.~\ref{fig:Multipoles}(b), is described by a nonzero quadrupole moment
$Q^{\alpha\beta} =  \sum_i h_i r_i^\alpha r_i^\beta$.
A solution of the Laplace's equation (\ref{DeltaMxy}) with the quadrupole symmetry behaves at large distance 
as \cite{LL2}
\begin{equation}
\phi({\bm r}) \propto\frac{\textrm{Re}\,(\hat{x}+i\hat{y})^2}{r^{D}} \,,
\label{Qab}
\end{equation}
which is slower than for the TAF.  Spins along the nodal $[11]$ and $[1\bar{1}]$ lines of the quadrupole moment 
passing through the impurity do not tilt irrespective of the distance to the vacancy as dictated by the four-fold 
rotation symmetry of the emerging spin texture. The asymptotic law (\ref{Qab}) suggests the same
behavior for spins  on the same lines that are off-centered from the vacancy as well.
 This is, however, not quite correct. Instead, spin tilting along the off-centered  $[11]$ and $[1\bar{1}]$ lines 
 is nonzero but follows the  $1/r^3$ distance dependence determined by the next-order multipole
 moment in the expansion of the source term (\ref{Qsource}).

The next two examples illustrated in Figs.~\ref{fig:Multipoles}(c) and \ref{fig:Multipoles}(d)
correspond to the $J_1$--$J_2$ Heisenberg antiferromagnet on a kagome lattice (KAF). The nearest-neighbor 
Heisenberg antiferromagnet with $J_1>0$ and $J_2=0$ has an extensive degeneracy of 
the classical ground states. For any triangular plaquette, any three spins are constrained 
to form a 120$^\circ$ configuration. However, this does not fix a periodicity of the magnetic structure
\cite{Chalker92,Harris92,Ritchey93}.
The second-neighbor exchange $J_2$ lifts this macroscopic degeneracy in two different 
ways depending on its sign. For $J_2>0$,  the $q=0$  state becomes energetically stable,
Fig.~\ref{fig:Multipoles}(c), whereas $J_2<0$ selects the $\sqrt{3}\times\sqrt{3}$
magnetic structure, Fig.~\ref{fig:Multipoles}(d), \cite{Harris92}. Having different 
translational patterns, the two states also differ in  the sense of spin rotations 
described by their chirality. The $q=0$ state exhibits the same chirality for all 
triangular plaquettes, whereas in the $\sqrt{3}\times\sqrt{3}$ structure 
the chiralities for up and down plaquettes have opposite signs.

A vacancy produces local strains with different symmetry for
the two magnetic  structures of KAF. This can be traced back to the presence of 
$\sin\theta_{ij}$ in the expression for the canting field (\ref{E1}), which in turn depends on the chirality 
pattern. For the $q=0$ state, the perturbation around a vacancy has a quadrupolar 
symmetry, Fig.~\ref{fig:Multipoles}(c), whereas for the $\sqrt{3}\times\sqrt{3}$ state 
the perturbation is of the dipolar type, Fig.~\ref{fig:Multipoles}(d). Hence, 
long-distance asymptotes are different for the vacancy-induced spin textures
in the two states: $1/r^2$ for 
the $q=0$ magnetic structure and $1/r$ for the $\sqrt{3}\times\sqrt{3}$ state. 
The horizontal [10] direction of the kagome lattice is parallel to the nodal line of the quadrupole moment
for the spin texture in Fig.~\ref{fig:Multipoles}(c).  Then, similar to the previous discussion for the FSAM model,
spins in the horizontal [10] chains  of a kagome lattice, Fig.~\ref{fig:Multipoles}(c),
readjust, though, with a faster $1/r^3$ decay law, which is determined by the subleading octupole term
in the multipole expansion of the canting field source  (\ref{Qsource}).

The slow $1/r$ decay of strain in a defect-induced spin texture is also characteristic for incommensurate 
helical structures. An example of such a system is an orthorhombically distorted TAF with different strength of 
exchange interactions in the horizontal chains ($J$) and on the interchain zigzag bonds ($J'$). 
A pattern of the canting fields produced by 
a vacancy still corresponds to Fig.~\ref{fig:Multipoles}(a), though the amplitudes 
$h_i^x$ become different for sites connected to an impurity by $J$ and $J'$ bonds. 
The source term, Eq.~(\ref{Qsource}),  has  in such a case a  
dipole symmetry. As a result,  magnetic distortion decreases
with distance as $1/r$, {\it i.e.}, significantly slower than the $1/r^3$ dependence that holds
 for the TAF with the six-fold rotation symmetry.

We conclude this section with a brief remark on the role of a magnetic anisotropy. 
If  the spin Hamiltonian (\ref{H}) contains, for example,  an additional easy-plane term, 
a noncollinear  magnetic state produced by the exchange interactions will select this plane.
The small-$k$ expansion for the out of plane distortion $m_i^y$ for a planar magnetic structure
contains now a finite gap in the denominator of Eq.~(\ref{Mya}). 
Accordingly, the long-distance behavior of the strain will change to an exponential decay with $r$
irrespective of the symmetry of the introduced perturbation.
The in-plane spin distortion $m_i^x$ has no such gap and, consequently, preserves its algebraic 
dependence with distance exhibiting a nonlocal behavior.

\subsection{Numerical results}
\label{CoplanarN}

Numerical minimization  of the classical energy (\ref{H})
in the presence of vacancies and other defects was performed for finite lattices using the periodic boundary conditions. The linear sizes $L$ of simulated
clusters are up to $L_{\rm max}=10^3$ unit cells  for two-dimensional (2D) spin models and up to $L_{\rm max}=10^2$ in three dimensions (3D).
We  have employed a standard iterative procedure,  which consists in sequential rotations of classical spins towards 
instantaneous local fields until a stationary state is reached, see
{\it e.g.}\  \cite{Gawiec91,Maryasin13,Wollny11}. For each simulated spin model,
we begin with a few  intermediate cluster sizes $L<L_{\rm max}$ and initialize the iteration process
with a random spin configuration and record  both the final stationary configuration
and its energy. The same procedure was repeated for up to 20--50 independent initial configurations
selecting in the end the lowest energy solution.  At the next stage a similar iteration procedure has been
performed starting with a fully ordered spin configuration that minimizes (\ref{H}) in the absence of  defects.
A comparison of final energies is made to ensure that the same fully relaxed spin structure is obtained
for both types of initial configurations. Numerical results for the largest lattices have been produced  
using only the fully ordered initial state.  Numerical data are shown for distances $r<L/2$, where the finite-size effects 
play no role. All distances are given in units of a spacing between the nearest-neighbor spins.

\begin{figure}[t]
\centerline{
\includegraphics[width=0.8\columnwidth]{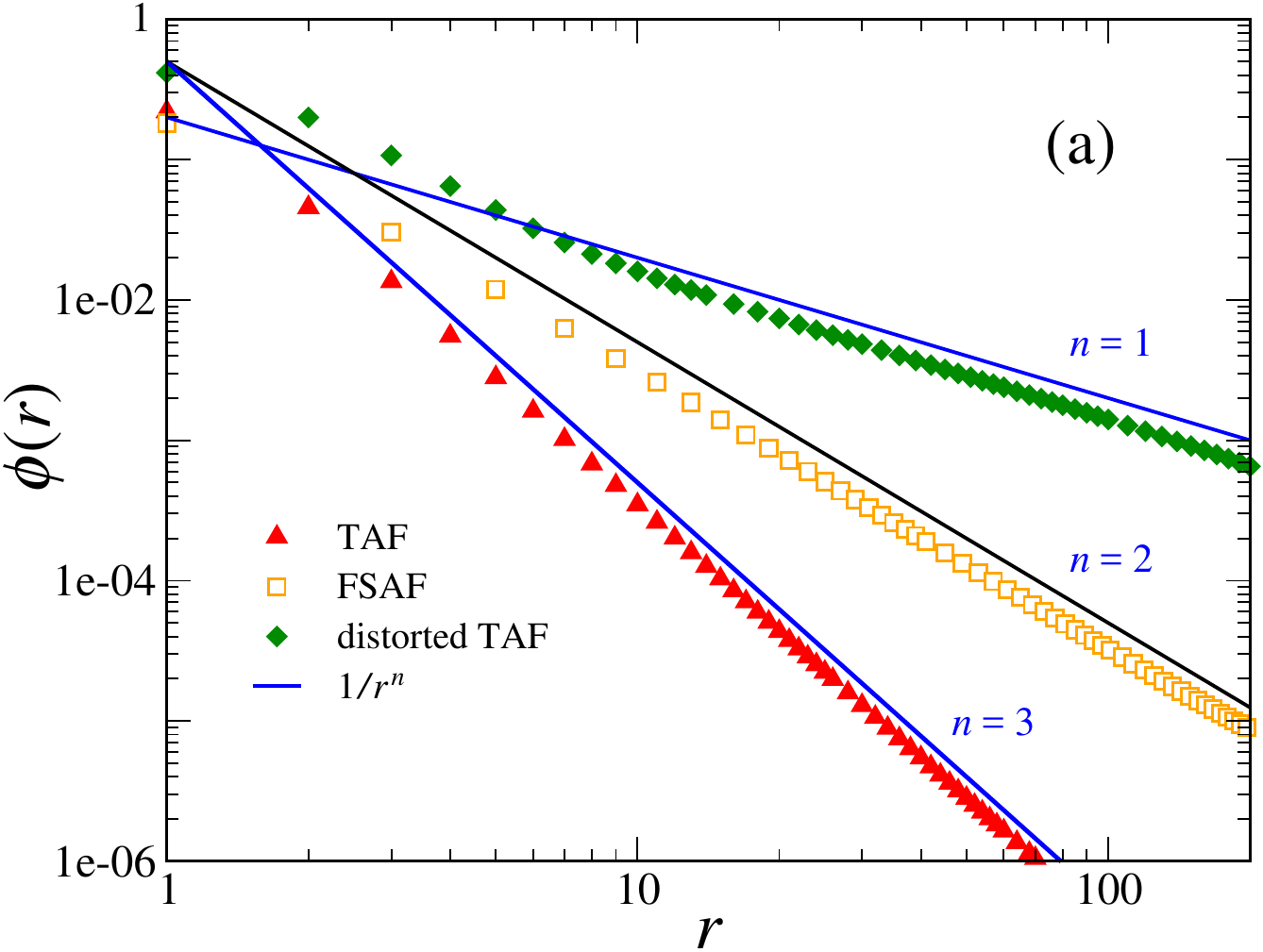}
}
\vskip 4mm
\centerline{
\includegraphics[width=0.8\columnwidth]{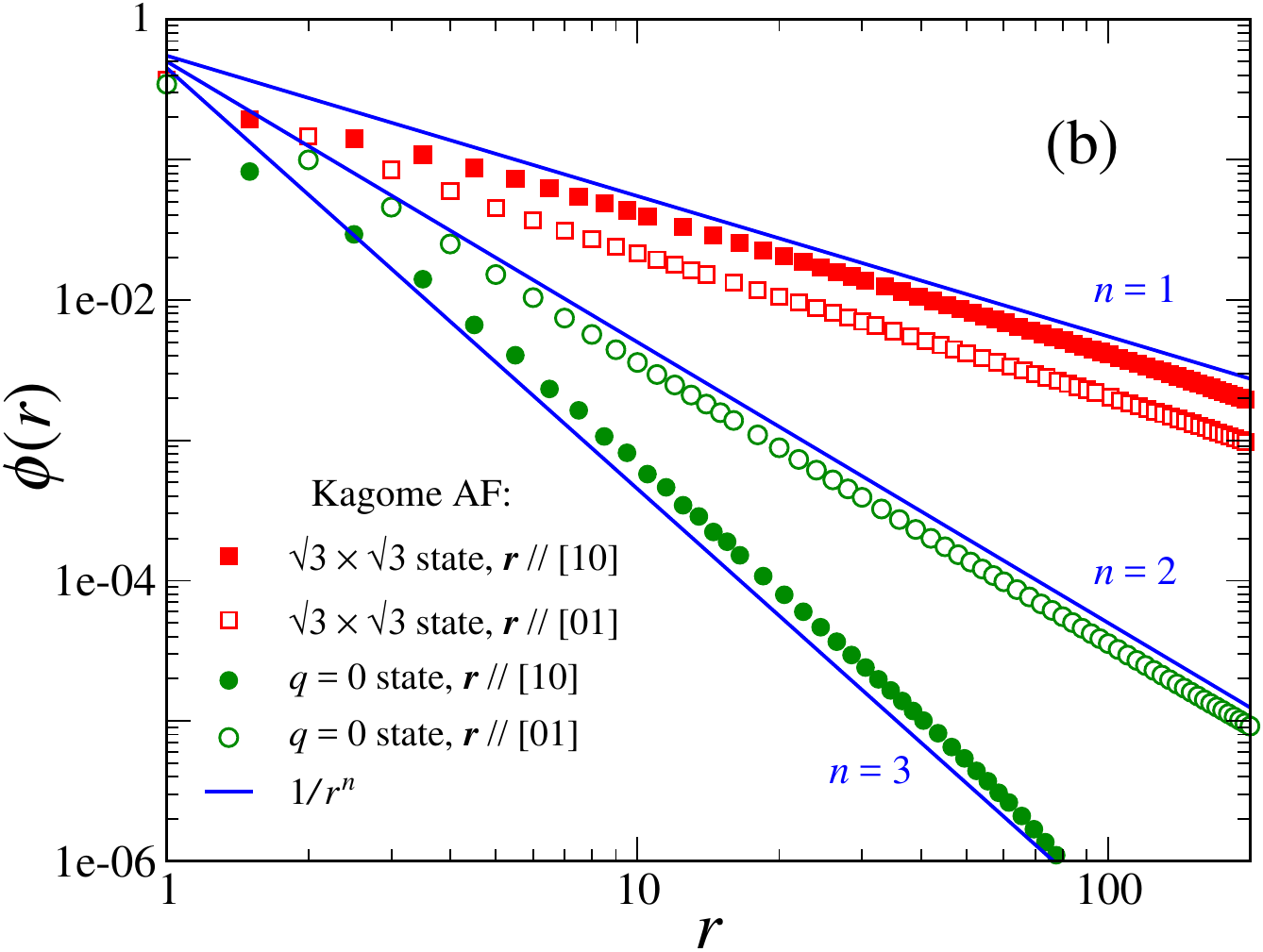}
}
\vskip 4mm
\centerline{
\includegraphics[width=0.8\columnwidth]{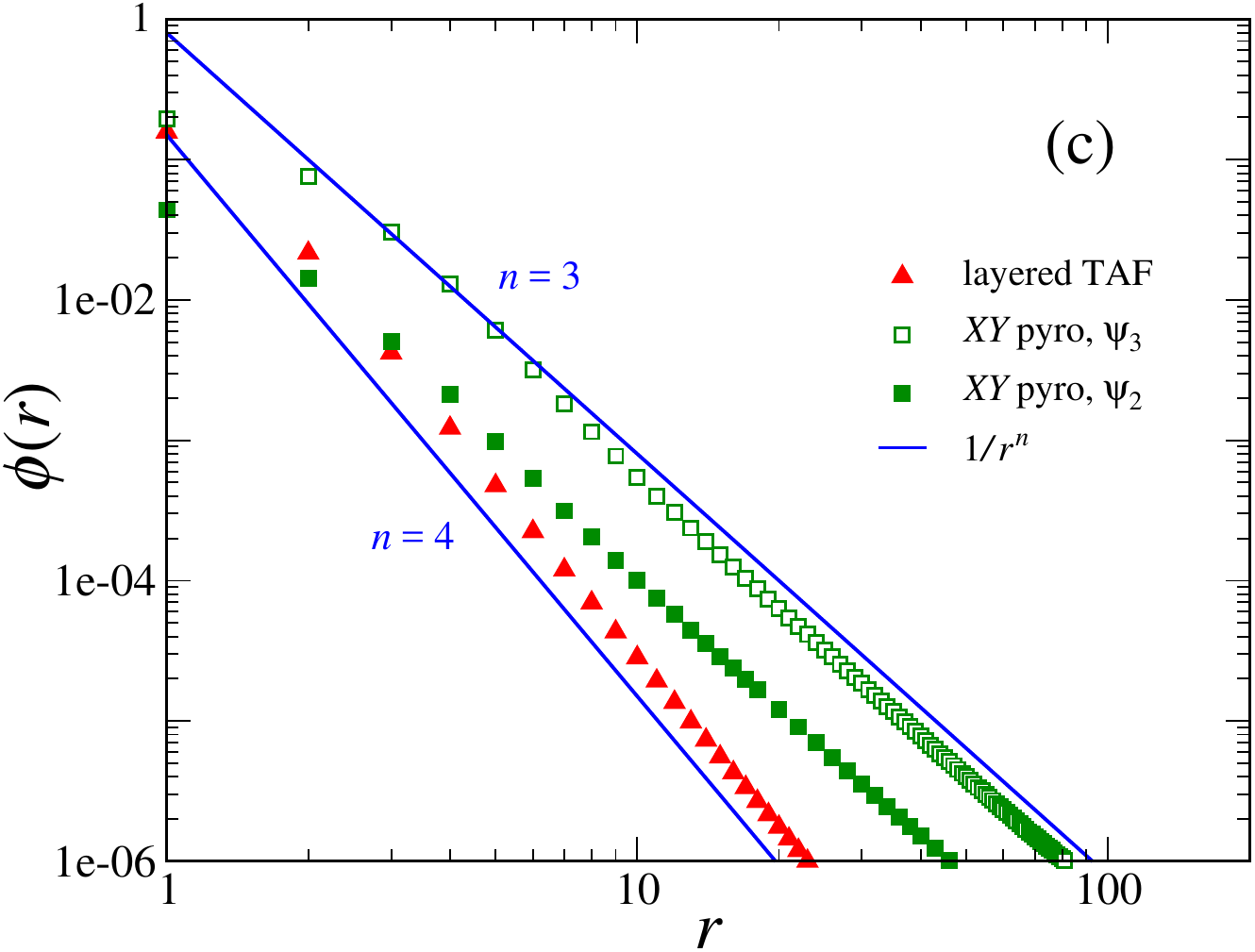}
}
\caption{Spin tilting angles $\phi(r)$ for a vacancy in coplanar magnetic states. 2D models: (a)
the triangular antiferromagnet (TAF), the frustrated square-lattice antiferromagnet (FSAFM),  and 
the orthorhombically distorted TAF; 
(b) the $J_1$--$J_2$ kagome antiferromagnet. 3D models: (c) the layered TAF  and the $XY$ pyrochlore antiferromagnets. 
The data points correspond to tilting angles in a relaxed state obtained numerically for a specific
ordered spin structure in the presence of a lattice defect.
}
\label{fig:asymptotic}
\end{figure}

Figure \ref{fig:asymptotic}(a) compares the decay of vacancy-induced spin distortions 
for three two-dimensional frustrated Heisenberg antiferromangets: on a  triangular lattice, on a square lattice 
with a diagonal exchange $J_2 = J_1$, and on a distorted triangular 
lattice with  $J' = 0.518J$, which corresponds to  an incommensurate spiral described 
by the wave vector ${\bm Q} = (7\pi/6,0)$. The deviation angle $\phi(r)$ is shown for
spins on the radial chains passing through the vacancy and parallel to the $[10]$ crystallographic direction, 
see Fig.~\ref{fig:Multipoles}. For FSAF, the tilting angles have 
different magnitudes on the two interpenetrating $\sqrt{2}\times\sqrt{2}$ sublattices and 
we show results only for one of them. For both sublattices the power law decay occurs
with the exponent $n=2$. In TAF, $\phi(r)$ decreases 
considerably faster with the power law exponent $n=3$. However, once the triangular lattice 
is orthorhombically distorted, the spin texture extends to much larger distances
and the power law exponent changes to $n=1$.

Next we consider the $J_1$--$J_2$ Heisenberg antiferromagnet on a kagome lattice
fixing $J_2/J_1 = \pm 0.1$ to stabilize either
the $q=0$ or the $\sqrt{3}\times\sqrt{3}$  structures, respectively. 
A vacancy in the $\sqrt{3}\times\sqrt{3}$ state produces the deformation
field with the dipole symmetry. Numerical results in Fig.~\ref{fig:asymptotic}(b) illustrate 
the corresponding $1/r$ decay for spin deviations along the $[01]$ chain, which
goes through the vacancy site, and along the horizontal $[10]$  chain, which passes 
through the base of the triangular plaquette 
containing the vacancy, see Fig.~\ref{fig:Multipoles}(d). 
For both chain directions the power law exponent is the same $n=1$.

Similar results for the $q=0$ state show a faster power-law decay of spin deviations with 
distinct exponents $n=2$ and 3 for the two chosen lines (chains).  The $n=2$ exponent  for the [01] chain
fully agrees with the quadrupole deformation field (\ref{Qab}).  On the other hand, the horizontal [10] chain
corresponds to a nodal direction of the quadrupole moment. Hence,  as was argued in Sec.~IIA,
the spin canting angle along this line is determined at large distances by the subleading octupole term.
The different vacancy response of the two coplanar 120$^\circ$ states of
the kagome antiferromagnet  may lead to their strikingly different stability  under the
doping. Differences in the  behavior  of the two coplanar states of the $J_1$--$J_2$  KAF at finite impurity
concentrations deserve further investigation.

The bottom panel in Fig.~\ref{fig:asymptotic}(c) shows numerical results for a selected three-dimensional 
spin models. The first example is the layered TAF with an interlayer coupling chosen as $J_c=J$ for 
an illustration. A pattern of transverse fields around a vacancy  remains the same as in 2D, Fig.~\ref{fig:Multipoles}(a),
irrespective of the sign of interplane coupling. Hence, an asymptotic form of the strain field is still given by 
the octupolar field Eq.~(\ref{Tabc}) with the $n=4$ exponent. Numerical results  for the tilting angles are  taken  
to a spin chain along the [100] direction. They are in full agreement with the analytic  asymptote.

The second 3D model is a pyrochlore antiferromagnet
with magnetic moments restricted to the $XY$ planes, which are orthogonal to the local 
trigonal axes. This model has been widely investigated in relation
to Er$_2$Ti$_4$O$_7$ and other rare-earth pyrochlores. We  refer 
the interested reader to \cite{Maryasin14,Zhitomirsky12} for further discussion and notations of this interesting model.
Depending on the ratio of two microscopic coupling constants, 
$J_\perp$ and $J_\perp^a$, the quantum fluctuations choose
either the noncoplanar state $\psi_2$ ($J_\perp^a/J_\perp<4$) or  
the coplanar structure $\psi_3$ as the ground state ( $J_\perp^a/J_\perp>4$) \cite{Zhitomirsky12}, 
whereas the structural disorder in the form vacancies or bond defects  has an opposite effect \cite{Maryasin14}.

The developed theoretical analysis of spin textures for the planar magnetic states applies here as well. Indeed, a strain 
for the $XY$ spins is described by a scalar field  $\phi({\bf r}_i)$, which represents a tilting angle in the local easy plane. 
Furthermore,  simultaneous rotation of  spins in the local planes between the  $\psi_2$ and  $\psi_3$
states costs no energy leading to a pseudo-Goldstone mode in the classical response.
Figure~\ref{fig:asymptotic}(c) shows a decay of spin distortion 
on a chain of spins passing through the vacancy in the
$\psi_2$ state  ($J_\perp^a/J_\perp=6$) and in the $\psi_3$ state ($J_\perp^a/J_\perp=1$).
The numerical results are compatible with  
the power-law exponent $n=3$ in both cases. Hence, the 
transverse field source term  must has the quadrupole symmetry (\ref{Qab}) for both states.
Calculation of  the  local canting fields according to Eq.~(\ref{Htrx})  confirms this conclusion.

\section{Noncoplanar spin textures}
\label{Noncoplanar}

\subsection{Analytic Theory}
\label{NoncoplanarA} 
 
The preceding analysis can be straightforwardly generalized to noncoplanar magnetic structures. 
Such spin arrangements can appear in various frustrated spin models \cite{Messio11,Zhitomirsky22}, 
especially in the presence of structural disorder, 
which tends to stabilize the least collinear states \cite{Henley89,Maryasin15}. 
Specifically, we consider Heisenberg nearest-neighbor antiferromagnets 
on a face-centered cubic and a pyrochlore lattices. 

A change in the orientation of an individual spin ${\bm S}_i$ from the reference state
${\bm S}_i^0$ can be described as 
a rotation about  an axis ${\bm n}_i$ ($|{\bm n}_i|=1$) by an angle $\phi_i$:
\begin{equation}
{\bm S}_i = {\bm S}^0_i\cos\phi_i + \bm{n}_i \times {\bm S}^0_i + \bm{n}_i (\bm{n}_i\cdot\bm{S}^0_i)(1-\cos\phi_i) \,.
\label{Rodrig}
\end{equation}
Assuming that $\phi_i$ is small and introducing $\bphi_i = \phi_i{\bm n}_i$ we obtain approximately
\begin{equation}
{\bm S}_i \approx {\bm S}^0_i + \bphi_i \times {\bm S}^0_i + \frac{1}{2}\, \bphi_i \times (\bphi_i \times{\bm S}^0_i) \,.
\label{Rodrig2}
\end{equation}
In the local frame with $\hat{\bm z}_i\parallel\bm{S}_i^0$, the above parameterization of a spin distortion  maps to  
 the spin tilting representation used in Sec.~\ref{CoplanarA} via 
\begin{equation}
 \bigl(\phi^x_i,\phi^y_i\bigr) \to  \bigl(-m_i^y,m^x_i \bigr) \,.
 \label{phiM}
\end{equation}
 The  value of $\phi_i^z$
does not change ${\bm S}_i$, since rotations about the spin itself are redundant.
 In particular, one can select a transverse gauge: $\bphi_i \cdot \bm{S}^0_i=0$.

Substituting (\ref{Rodrig2})  into the spin Hamiltonian (\ref{H})  we obtain expansion of the classical energy in powers
of spin distortion. The linear terms are given by
\begin{equation}
E_1   =  -\sum_i \bm{\phi}_i \cdot \bm{\tau}_i \,, \quad \bm{\tau}_i = -\sum_j \delta \tilde{J}_{ij} 
\bigl({\bm S}^0_i \times {\bm S}^0_j\bigr)\,,
\end{equation}
which is an analog of Eq.~(\ref{E1}). Because of Eq.~(\ref{phiM}),
the spin distortion field $\bm{\phi}_i$ couples linearly to a torque $\bm{\tau}_i$ 
rather than to a transverse field $\bm{h}_i$. 

The deformation energy  (to quadratic order in $\bphi_i$) in  the noncoplanar case is given by
\begin{eqnarray}
E_2  & = &\frac{1}{2}\!  \sum_{\langle ij\rangle} J_{ij}   \Bigl[- \bigl({\bm S}^0_i\! \cdot {\bm S}^0_j\bigr) (\bphi_i\! - \bphi_j)^2 
+ (\bphi_i\! \cdot {\bm S}^0_i) (\bphi_i\! \cdot {\bm S}^0_j) 
\nonumber 
\\ 
&  & \mbox{} +  (\bphi_j \cdot {\bm S}^0_i) (\bphi_j \cdot {\bm S}^0_j) 
- 2(\bphi_i \cdot {\bm S}^0_j) (\bphi_j \cdot {\bm S}^0_i)  \Bigr]
\label{eqn:NCSmallDefH}
\end{eqnarray}
Varying the total energy $E_1 + E_2$ with respect to $\bphi_i$, we obtain
\begin{eqnarray}
&& \sum_j J_{ij} \Bigl\{2({\bm S}^0_i \cdot {\bm S}^0_j) (\bphi_j - \bphi_i) 
 + \bigl[{\bm S}^0_i \cdot (\bphi_i - 2\bphi_j)\bigr] {\bm S}^0_j  
 \nonumber  \\ 
&& \mbox{\qquad\qquad\qquad\qquad}  + (\bphi_i \cdot {\bm S}^0_j) {\bm S}^0_i \Bigr\} = {\bm \tau}_i 
\label{eqn:NCspindef}
\end{eqnarray}
used below in numerical calculations of the spin deformation field $\bphi_i$.

To transform to the continuum limit, we assume that $\bphi_i$ is a slowly varying function of distance $\bphi({\bm r})$
and choose the transverse gauge. Furthermore, we approximate
\begin{equation}
\bphi_i  - \bphi_j \approx (\bm{r}_{ij}\cdot\bm{\nabla})\bphi(\bm{r})\,.
\end{equation}
Then, the continuum form of the deformation energy is
\begin{equation}
\label{eqn:NCElas}
E_2 = \frac{1}{2} \int {\rm d}^D{\bm r}\; K_{\alpha \beta}^{ab} \,(\nabla_\alpha \phi^a)(\nabla_\beta \phi^b)\,.
\end{equation}
Here, $a,b$ are indices in the spin space and $\alpha,\beta$ denote directions in the real space.
In comparison to a similar expression for the coplanar case (\ref{Egrad}), the elastic constant becomes
 a forth-rank tensor determined by
\begin{eqnarray}
K^{ab}_{\alpha \beta} & = & - \frac{1}{V_0}\sum_{j} J_{ij} \,
{r}_{ij}^\alpha {r}_{ij}^\beta
\Bigl[ ({\bm S}^0_i \cdot {\bm S}^0_j ) \delta^{ab} \\
&  & \mbox{\qquad\qquad} -  (S^0_i)^a (S^0_j)^b -  (S^0_j)^a (S^0_i)^b)  \Bigr]
\nonumber
\end{eqnarray}
where $V_0$ is the volume of the unitcell.
For a noncoplanar magnetic structure, different components $\phi^a$ are coupled to each other
by the non-diagonal terms of the elastic tensor $K^{ab}_{\alpha \beta}$.

The presence of a vacancy in this system can be modeled as a source term in Eq.~(\ref{eqn:NCElas}).  The large-distance behavior of the spin deformations about a vacancy can again be predicted from the properties of Green's function of the elasticity theory. Thus, for a delta function source located at the origin
\begin{equation}
|\bphi(r)| \sim \frac{1}{r^{D-2}} \,,
\end{equation}
where $D$ is the dimensionality of space.
For a source torque described by the lowest multipole $M$, the lond-distance behavior  changes to
\begin{equation}
\label{eqn:MultiPoleLongDistance}
|\bphi(r)| \sim \frac{1}{r^{D+M-2}}
\end{equation}
If the source has higher multiples than $M$, then one may observe a faster decay along special directions where the contributions of the lowest multipole vanishes owing to symmetry similar to the discussion in Sec.~IIA.

\begin{figure}[t]
\includegraphics[width=0.85\linewidth]{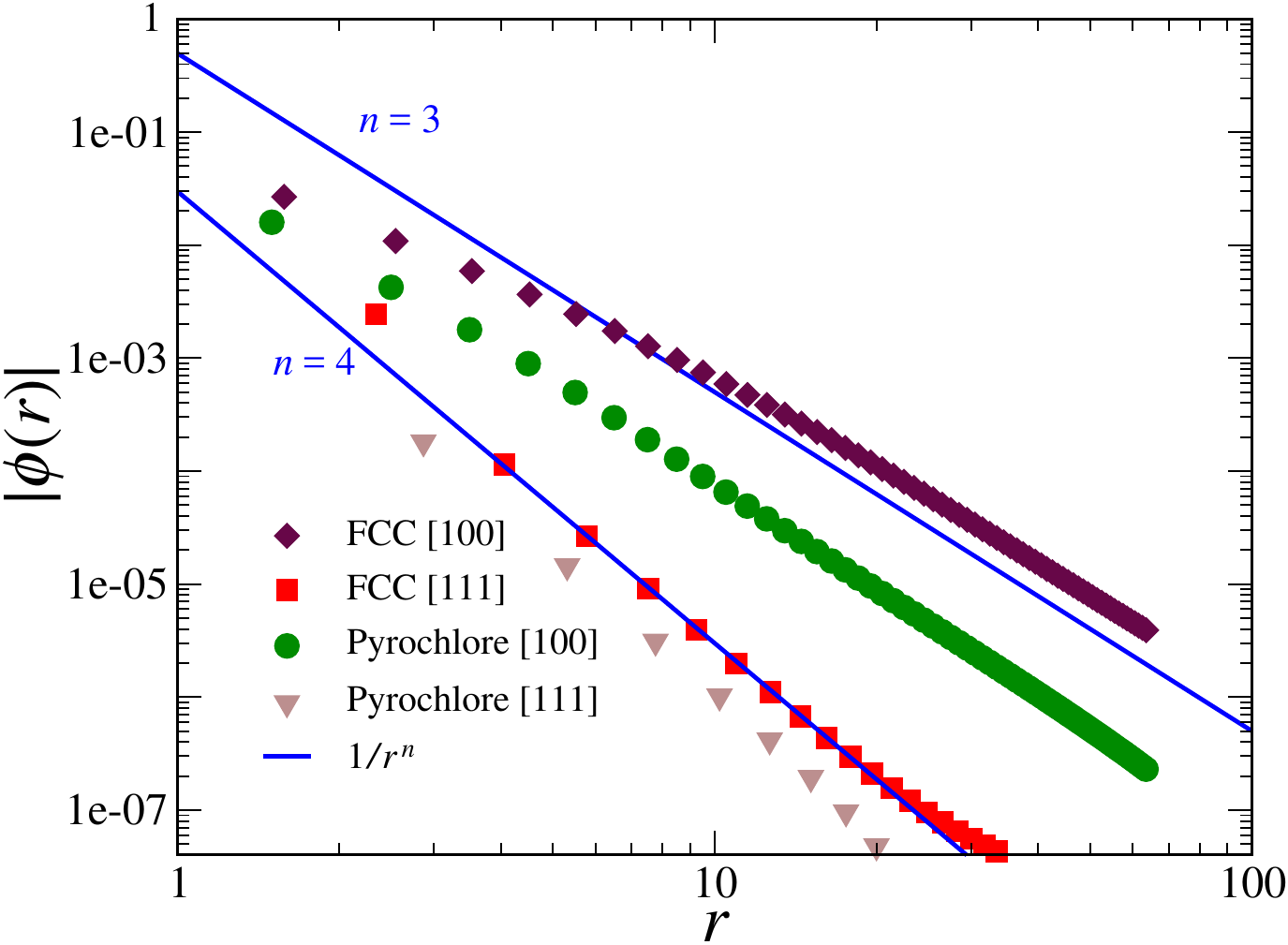}
\caption{
Distortion angle $\phi$ versus distance around a vacancy in fcc and pyrochlore antiferromagnets with the all-in-all-out 
noncoplanar states. Numerical results are obtained by minimization
of energy (\ref{eqn:NCSmallDefH}) using a periodic box with the linear size $L=256$.
[100] and [111] are the chosen crystallographic directions. }
\label{fig:NonCoplanar} 
\end{figure}

\subsection{Numerical results}
\label{NoncoplanarN}

The above theory is verified for the two classical models with noncoplanar magnetic structures:
the nearest-neighbor Heisenberg antiferromagnets on  fcc and pyrochlore lattices.  The starting point
is the linear equation on  $\bphi_i$ (\ref{eqn:NCspindef})
supplied with a torque term $\bm{\tau}_i$ produced by a vacancy on the neighboring sites.
This simpler linear approach is adopted in three-dimensional cases as the systems sizes required to obtain the large-distance behavior of spin-deformation are prohibitive to be treated by the full minimization of energy as adopted in the two-dimensional 
systems of the previous section.  Here, we numerically solve the linear equations (\ref{eqn:NCspindef})  
with the Fourier transform method, first computing $\bphi(\bm{k})$
followed by an inverse transformation to obtain $\bphi_i$. 
The calculations have been performed on periodic lattices with $L^3$ unit cells with $L \leq 256$.

\begin{figure}[b]
\centerline{\includegraphics[width=0.6\linewidth]{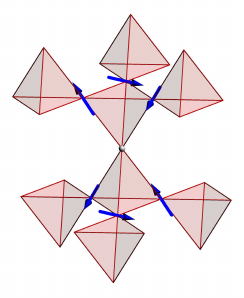}}
\caption{\label{fig:PYCTorque} Distribution source terms due to a vacancy in the pyrocholore antiferromagnet. 
The vacancy is indicated by a missing site, and the blue vectors indicate the source terms (uncompensated local magnetic field crossed with the local spin) induced by the missing site. The lowest vanishing multipole moment is the quadrupole moment.}	
\end{figure}

Figure~\ref{fig:NonCoplanar} shows the behavior of spin deformation fields in fcc and pyrochlore antiferromagnets induced by a vacancy. 
The ground state is the all-out state in both cases. The pyrochlore case has a large ground state degeneracy \cite{Moessner1998}. 
The spin deformation in this case is defined around the all out state, with possible zero energy local deformations. We find that along a generic direction, 
the spin deformation field decays as $1/r^3$. Further, along special directions (e.g., the [111] direction) we see a faster decay of $1/r^4$. 
These results are readily understood from the theory developed earlier. The vacancy induces source terms on the neighbor sites. Figure~\ref{fig:PYCTorque} 
shows the distribution of these source terms near a vacancy. In this case (and also for the fcc), we see that the lowest nonvanishing multipole moment 
of this source term is the quadrupole moment, i.e., the long distance  behavior from Eq.~(\ref{eqn:MultiPoleLongDistance}) is indeed $1/r^3$ as found in numerical 
calculation. Further, the quadrupole field vanishes along certain directions; along these directions the octupole moment of the source term produces 
an $M=3$, i.e., a $1/r^4$ decay of the spin deformation.

\section{Defect induced moments}
\label{Moments}

A removal of one spin from an antiferromagnetic crystal leaves an uncompensated magnetic 
moment. For noncollinear magnetic structures the resulting net moment is additionally screened 
by the induced spin texture \cite{Wollny11}. Assuming that an impurity is located at $i=1$, 
the total moment can be written as
\begin{equation}
{\bm m}_{\rm imp} = \sum_{i\neq 1} {\bm S}_i =  \sum_{i\neq 1} {\bm m}_i - {\bm S}^{0}_1 \ .
\label{Mvac}
\end{equation}
The last form is independent of the boundary conditions and can be used for commensurate 
as well as for incommensurate states. Generally, the moment $m_{\rm imp}$
has a nonuniversal fractional value, which  has to be determined numerically. 
The direction of ${\bm m}_{\rm imp}$ is, however, fixed by the background
magnetic structure and in the absence of further symmetry breaking is locked to the
direction of the missing spin ${\bm S}^{0}_1$.

The vacancy moments manifest themselves as the low-temperature Curie tail in the magnetic susceptibility
\begin{equation}
\chi_{\rm imp} \simeq \frac{N_{\rm imp}m_{\rm imp}^2}{3T}\ ,
\label{chi}
\end{equation}
where $N_{\rm imp}$ is a number of nonmagnetic impurities in the crystal
\cite{Wollny11,Maryasin15}. 
The effect is present for both 2D and 3D magnets irrespective of an ordering transition
in the later case. Despite apparent similarity with the result for  a paramagnet consisting of
$N_{\rm imp}$ independent moments, Eq.~(\ref{chi}) has a somewhat different meaning. 
As we explained before, the vacancy moments are locked to the specific sublattice directions 
in the ordered  magnetic structure and do not fluctuate. Since random vacancies
appear with equal probability on each of the antiferromagnetic sublattices, 
their total moment vanishes $\langle M_{\rm imp}\rangle=0$ for $N_{\rm imp}\gg 1$.
However, the average square moment remains nonzero and scales as 
$\langle M_{\rm imp}^2\rangle \approx N_{\rm imp}m_{\rm imp}^2$. It is this  
moment induced by configurational fluctuations which contributes to the low-temperature 
singularity in the magnetic susceptibility (\ref{chi}).

Let us now consider how the different decay laws for spin textures
affect the vacancy moment (\ref{Mvac}). In particular, the important question is
if the dipole-like deformation field gives finite or diverging contributions
from large distances. For concreteness, we consider a distorted incommensurate spiral
\begin{equation}
{\bm S}_i =\bigl(\sin({\bm Q}\cdot{\bm r}_i+\phi_i), 0 ,\cos({\bm Q}\cdot{\bm r}_i+\phi_i)\bigr)
\end{equation}
with $\phi_i$ obeying Eq.~(\ref{Dipole}). Then, the magnetic moment of the spin texture
is approximately given by
\begin{equation}
\delta m_{\rm imp}^z = - \sum_i  \frac{{\bm d}\cdot{\bm r}_i}{r_i^D}\, \sin({\bm Q}\cdot{\bm r}_i) \,.
\label{Mdip0}
\end{equation}
Replacing the lattice sum  by a space integral we obtain in 2D for the contribution
accumulated at large distances $r>a^*$:
\begin{eqnarray}
&& \delta m_{\rm imp}^z = - d \iint \frac{\textrm{d}x\textrm{d}y}{V_0}\, 
\frac{x\sin Qx}{x^2+y^2}  
\nonumber \\
&&  \mbox{\ \ } = - d\int_{a^*}^\infty\frac{\textrm{d}r}{V_0}
\int_0^{2\pi}\textrm{d}\varphi\, \cos\varphi\,\sin(Qr\cos\varphi)
\nonumber \\
& & \mbox{\ \ } = - 2\pi d\int_{a^*}^\infty\frac{\textrm{d}r}{V_0}\, J_1(Qr) =
 -\frac{2\pi d}{V_0Q}\, J_0(Qa^*) \ ,
\label{Mdip}
\end{eqnarray}
where $d$ is the dipole moment of the spin distortion
field,  $V_0$ is the unit cell volume, 
and $J_0(x)$, $J_1(x)$ are the two Bessel functions.
Thus, the long-range contribution to $m_{\rm imp}$ remains finite.
This conclusion can be easily extended to 3D and to other types of the 
magnetic deformation fields (\ref{Tabc}) and (\ref{Qab}).

\begin{table}
\caption{\label{parameters} 
Magnetic moments induced by vacancies in various 2D and 3D models. Defect (impurity) bond
results with $J_{\rm imp}\neq J$ are also included for TAF.
 }
\begin{center}
\begin{tabular}{lclcrcl}
\hline
\hline
 Model && Parameters && $M_{\rm imp}/S$ & \mbox{$\quad$} & Power law\\
\hline
& & \\
TAF, vacancy  && && $-0.03974$ && $\quad 1/r^3$ \\
TAF, bond     && $J_{\rm imp} = 0.5J^{^{}}$ && 0.19383 && $\quad 1/r$  \\
              && $J_{\rm imp} =-0.5J^{^{}}$ && 0.52928 && $\quad 1/r$  \\
\hline
FSAF && $J_2/J_1 = 1$   &&  0.50289    && $\quad 1/r^2$ \\
FSAF && $J_2/J_1 = 0.7$ &&  0.29232    && $\quad 1/r^2$ \\
\hline
KAF  && $J_2/J_1 =0.1$  &&  $-0.04242$ && $\quad 1/r^2$ \\
     && $J_2/J_1 =-0.1$ &&  $-0.02719$ && $\quad 1/r$ \\
\hline
distorted TAF && $J'/J =0.765$ && 0.04398 && $\quad 1/r$ \\
              && $J'/J =0.518$ && 0.25827 && $\quad 1/r$ \\
\hline
layered TAF   && $J_c/J = 1$    && 0.43213 && $\quad 1/r^4$ \\
              && $J_c/J = 0.1$  && 0.03969 && $\quad 1/r^4_{_{}}$ \\              
\hline
\hline
\end{tabular}
\end{center}
\end{table}

Table I presents numerical values for the  vacancy moments in
a few selected spin models. Positive or negative value of the moment corresponds to
whether it is antiparallel or parallel to the missing spin ${\bm S}_1^{0}$. 
In a collinear magnetic state, the vacancy moment is equal to 
$-{\bm S}_1^{0}$ and, hence, is positive according to the above definition.
In TAF the missing spin is oversceened by the induced spin texture
such that the moment becomes negative \cite{Wollny11}. The sign change is also present 
for vacancies in KAF, where spin textures exhibit a different 
large-distance behavior from TAF. Hence, underscreening or overscreening 
is mostly produced by local spin canting in a few nearest-neighbor shells
around the impurity.

As discussed in Sec.~II, a defect bond also generates the dipole type
spin texture (\ref{Dipole}). Hence, a finite value of the lattice
sum in Eq.~(\ref{Mdip0}) signifies that defect bonds in noncollinear antiferromagnets
must also produce  uncompensated magnetic moments. 
Their direction is now fixed by the sum of two spins on each defect bond
and the fractional values can widely vary. To our knowledge 
this effect has not been so far discussed in the literature on structural defects.
For illustration, we compute and present in Table I the values of magnetic moments 
associated with a defect bond $J_{\rm imp}\neq J$  in TAF for two values of $J_{\rm imp}$.
Note, that the model of defect bonds in TAF is relevant to the triangular  
antiferromagnet RbFe(MoO$_4$)$_2$ for which nonmagnetic substitution of Rb by K 
introduces random modulations of exchange bonds between iron ions \cite{Smirnov17}.

\section{Discussion}
\label{Discussion}

We have presented a comprehensive theoretical  analysis of spin textures induced by lattice defects, in particular,
nonmagnetic vacancies, in ordered noncollinear magnets. {Our results explain the apparently conflicting results in 
the literature  about the exponent of the power-law tail in the distance dependence of vacancy induced spin distortions 
\cite{Henley01,Wollny11,Weber12}. Besides space dimension, the corresponding power-law exponent is determined,
by symmetry of the canting field source, which in turn depends on the magnetic structure.
Perhaps the most striking example of this conclusion is provided by the $q=0$ and the $\sqrt{3}\times\sqrt{3}$ states,
on a kagome lattice. Both states have locally the same $120^\circ$ magnetic structure albeit with different 
arrangements of spin chiralities. Such a difference leads to  different decay laws of the spin distortions:
$1/r^2$ ($q=0$) and $1/r$ ($\sqrt{3}\times\sqrt{3}$).}  Another example, is the triangular lattice antiferromagnet with
the $1/r^3$ decay law for the vacancy-induced spin texture, which changes to the much slower $1/r$ dependence
in the presence of an arbitrary weak orthorhombic lattice distortion.

The developed elastic-type theory is inherently concerned with the long-wavelength physics. While input from the lattice scale is important and useful, such in the case of determining the multipolar nature of a defect mentioned above, our theory yields the long-distance asymptotics and not necessarily the full quantitative form of the distortion at short distances.

The above consideration can be straightforwardly extended to other types of structural defects 
and associated perturbations in the spin Hamiltonians. These may include magnetic impurities or add-on spins,
locally modified spin-orbit coupling as, for example, the Dzyaloshinskii-Moriya term \cite{Hayami19} and so on.
Similar to the problem of an antiferromagnetic bond in a collinear ferromagnet \cite{Villain79,Aharony88,Parker88,Vannimenus89}, 
in some of these cases an additional symmetry lowering can be expected
once the strength of a local perturbation exceeds a critical value. An interesting aspect that may deserve further numerical
studies  is the Friedel-like oscillations in the transverse response for the coplanar magnetic structures.

Our single-impurity results have important consequence for the interaction between  impurities, and thus their eventual behavior at finite impurity concentration. The elastic theory  implies long-range interactions mediated by the induced spin distortions. As a consequence, 
longer-range tails in the induced spin distortions should lead to fragile magnetic phases and faster transformation
into a disordered spin-glass like state upon doping with nonmagnetic impurities. It will be interesting to perform direct numerical studies of such effects. 

Finally, an interesting direction concerns interplay and competition between thermal and quantum fluctuations and the response to vacancies in the many classical ground states in frustrated magnets which are degenerate but not symmetry-related. While competing  tendencies of fluctuation- and disorder-induced ordering have been known for a long time \cite{Henley89}, we are not aware of a comprehensive study of all the possible combinations of unperturbed states and disorder types for determining the eventual low-temperature behavior of such magnets.

\acknowledgements
We thank M. Vojta for discussions. M.\,E.\,Z.\ acknow\-ledges support from the ANR, France, Grant No. ANR-19-CE30-0040. 
V.\,B.\,S.\ thanks SERB, DST, India for support.
This work was supported in part by
the Deutsche Forschungsgemeinschaft under Grant No.\ SFB
1143 (Project-ID No.\ 247310070) and by
the Deutsche Forschungsgemeinschaft under cluster of excellence
ct.qmat (EXC 2147, Project-ID No.\ 390858490).

\end{document}